\documentstyle[12pt,fleqn]{article}
\def\a{\alpha }   \def\b{\beta }
   
   \def\D{\Delta }  
   \def\L{\Lambda }  
   \def\O{\Omega }
   \def\S{\Sigma}
\def\i{{\rm i}}

\def\beq{\begin{equation}}
\def\eeq{\end{equation}}
\def\ra{\rangle}
\def\la{\langle}
   
\begin{document}
    \begin{flushright}
          {hep-ph/9807559}\hspace{14mm}\\
          {July 1998}\hspace{14mm}\\
    \end{flushright}

\vspace{18mm}
\begin{center}
\large{\bf
Quantum Groups as Flavor Symmetries:
Account of Nonpolynomial $SU(3)$-Breaking Effects
in Baryon Masses
}
\end{center}

\vspace{12mm}
\begin{center}
{A.M.~Gavrilik\footnote{E-mail: omgavr@bitp.kiev.ua}, \  N.Z.~Iorgov}
\end{center}
\vspace{2mm}
\begin{center}
{\it Bogolyubov Institute for Theoretical Physics,  \\
     Metrologichna str. 14b, Kiev-143, Ukraine}
\end{center}
\medskip

\vspace{14mm}
\begin{abstract}
  The use, for flavor symmetries, of the quantum (or $q$-) analogs of
unitary groups $SU(n_f)$ yields new, very accurate, baryon mass sum
rules. We show, in the 3-flavor case, that such approach accounts for
nonpolynomial $SU(3)$-breaking effects in the octet and decuplet baryon
masses. A version of this approach with manifestly $q$-covariant mass
operator is given. The obtained new version of the $q$-deformed mass
relation is simpler than those derived before, but, for its empirical
validity, the parameter $q$ is to be fixed by fitting. As shown, the
well-known Gell-Mann--Okubo octet mass sum rule results, besides usual
$SU(3)$, also from an exotic "symmetry" encoded in the singular case
$q=-1$ of the $q$-algebra $U_q(su_3)$.
\end{abstract}


\newpage
\bigskip
{\bf 1.\ }
Application of quantum groups and quantum, or $q$-deformed, algebras
[1] to diverse physical problems remains the subject of intensive
study (see, e.g., overviews [2]). In recent years, some attempts
were made to exploit various deformed algebras in the context of hadron
phenomenology, both in the scattering sector [3] and in the sector
of such static properties as (ground state) hadron masses [4-8].
It was found, in particular, that $U_q(su_n)$-based approach
enables one to obtain mass sum rules of an accuracy essentially
improved not only with respect to the equal spacing rule (ESR)
$M_\O-M_{\Xi^*}=M_{\Xi^*}-M_{\S^*}=M_{\S^*}-M_{\D}$ for decuplet
baryons, but also in comparison to the slightly better
decuplet formula [9,10] $M_\O-M_{\D}=3(M_{\Xi^*}-M_{\S^*})$, and
even in comparison to the famous Gell-Mann--Okubo (GMO) octet
sum rule [11] $M_N+M_\Xi=\frac{3}{2}M_\L+\frac{1}{2}M_\S$
which holds with $0.58\%$ accuracy.

Our main purpose in this note is to demonstrate transparently
that the application of the $q$-algebras $U_q(su_n)$ taken for
flavor symmetries of hadron dynamics efficiently takes into account
the $SU(3)$-breaking effects which are of highly nonlinear nature,
namely, nonpolynomial effects. As another result, we obtain,
utilizing properties of $q$-tensor operators and
the important ingredients of the Hopf algebra structure
of $U_q(su_n)$ (comultiplication, antipode), a different version
of the $q$-deformed mass relation, see (29) below. The new
interesting feature of this relation is that it produces the classical
GMO sum rule not only in the classical limit $q=1$ but also in the
nonclassical situation of $q=-1$. A description of the mathematical
structure (operator algebras) corresponding to $q=-1$
is also given.$\!$\footnote
{The quantum algebra $U_q(su_n)$ is known to be undefined
in this case [1,12]. }

\bigskip
{\bf 2.\ }
In order to describe mass splittings
for particles from isomultiplets within the octet of baryons
$J^P =  \frac12 ^+$,
we adopt, like in [4-7], that the algebra of hadron flavor symmetry
in the 3-flavor case is not $su_3$, but its $q$-analogue $U_q(su_3)$
(broken down to the isospin algebra $U_q(su_2)$).
We use a correspondence between baryons of the octet and basis vectors
of the carrier space of irreducible representation (irrep) {\bf {8}}
of the highest weight $\{p+2,p+1,p\}$, $p \in {\bf Z}$,
of $U_q(su_3).$. Here and below, we will take the highest weight
of irrep of $U_q(su_n)$ as that of $U_q(u_n)$, thus identifying
two weights of the latter if their difference is a weight of the
form $\{s,s,\ldots,s\} , s \in {\bf Z}$ (our final results will be
insensitive
w.r.t. such a difference). Let us remark that necessary details
concerning the "compact" $U_q(u_n)$ and "noncompact" $U_q(u_{n,1})$
algebras as well as their irreps can be found, e.g., in [6].

Since the algebra $U_q(u_2)$
will be always unbroken within our treatment (exact
isospin--hypercharge symmetry of strong interaction),
we have mass degeneracy within each
isomultiplet of octet and, thus, 4 different masses within the octet.
Isomultiplets from the octet are put into correspondence with
the following representations $\{m_{12},m_{22}\}$ of $U_q(u_2)$:
\[
N \leftrightarrow \{p+2,p+1\} \ \ , \ \ \ \ \
\Sigma \leftrightarrow  \{p+2,p\} \ \ ,
\]
\[
\Lambda \leftrightarrow \{p+1,p+1\} \ \ , \ \ \ \ \ \
\Xi \leftrightarrow \{p+1,p\} \ \ ,
\]
and will be denoted as $| B_i \ra$, where $i$ runs over
isoplets of the octet, i.e. $i=1,...,4\ .$
The masses $M_{B_i}$ of particles are given as diagonal
matrix elements
\beq
M_{B_i}=\la B_i|\hat{M}|B_i\ra
\eeq
of the mass operator [6,5]
\beq
\hat{M}=\hat{M}_0 + \alpha \hat L + \beta \hat R,
\eeq
\[
\hspace {10mm} \hat{M}_0=\widetilde{M}_0 {\bf 1} +
\gamma A_{45} A_{54} + \delta A_{54} A_{45},
\]
\beq
\hspace {10mm} \hat L = A_{35} \tilde {A}_{53} +
\tilde{A}_{35} A_{53}\ \ , \ \ \ \
\hat R = A_{53} \tilde {A}_{35} +  \tilde{A}_{53} A_{35}
\eeq
which is composed from operators representing
basis elements of the $U_q(su_5)$ algebra
\footnote {
 We embed the $U_q(su_3)$ algebra
into $U_q(su_4)$ and further into $U_q(su_5)$,
in order to lift the $\Lambda$--$\Sigma$ mass degeneracy.
}
and acting in the space of "dynamical" representation
$\{m_{15},p+1,p,p,m_{55}\}$ of this algebra.

In accordance with the chain of embeddings of $q$-algebras
\[
U_q(su_3)\subset  U_q(su_4)\subset   U_q(su_5)
\]
there is the corresponding embedding of the representations
under consideration:
\beq
\{p+2,p+1,p\}\subset \{p+2,p+1,p,p\} \subset \{m_{15},p+1,p,p,m_{55}\}.
\eeq

It is clear from the action formulas of $A_{45}$ and $A_{54}$
(see Jimbo in [1], and [12]) that the operators
$A_{45}A_{54}$ and $A_{54}A_{45}$
do not give splitting between the isoplet masses $M_{B_i}$,
but only shifting the common background mass $\widetilde{M}_0.$
For that reason, we put $\la B_i|\hat{M}_0|B_i\ra=M_0$.

Diagonal matrix elements of the operators
$\hat L$ and $\hat R$ in (3) can be rewritten as
\footnote{
Symbol $[x]$ denotes the $q$-number
$[x]\equiv [x]_q\equiv (q^x-q^{-x})/(q-q^{-1})$
corresponding to the number $x$; the latter, conversely, results
as the 'classical' limit of $[x]$, that is, $[x]\to x$ if $q\to 1$.
}
\[
\langle B_i | ( A_{35} \tilde {A}_{53} +  \tilde{A}_{35} A_{53})
| B_i \rangle
\]
\beq
\hspace {6mm}
 = \langle B_i | ( [2] A_{34} A_{45} A_{54} A_{43} +
[2] A_{45} A_{34} A_{43} A_{54} - 4 A_{34} A_{45} A_{43} A_{54} )
| B_i \rangle,
\eeq
\[
\langle B_i | ( A_{53} \tilde {A}_{35} +  \tilde{A}_{53} A_{35} )
| B_i \rangle
\]
\[
\hspace {6mm} = \langle B_i | ( [2]  A_{54} A_{43} A_{34} A_{45}+
[2] A_{43} A_{54} A_{45} A_{34}  - 4  A_{54} A_{43} A_{45} A_{34} )
| B_i \rangle
\]
\beq
\hspace {6mm} = [2] \langle B_i | A_{54} A_{43} A_{34} A_{45}
| B_i \rangle.
\eeq
The last equality in (6) is a consequence of
the particular choice of representations in (4).
Matrix elements (5) and (6) are evaluated [6] in the framework of
the Gel'fand- Tsetlin formalism. Here, we take into account the
identifications (we put $p=0$)
\vspace{-1mm}
\[
m_{12}=\frac Y 2 +I+1  \ ,\ \ \ \ \  m_{22}=\frac Y 2 -I+1\ ,
\]
where $Y$ and $I$ are hypercharge and isospin, respectively
(their values label each isoplet unambiguously), and obtain
for the summands in (5) and (6) the following expressions:
\beq
\langle B_i | A_{45} A_{34} A_{43} A_{54} | B_i \rangle =
-[m_{15}+4][m_{55}][6]^{-1}[2-Y],
\eeq
\beq
\langle B_i | A_{34} A_{45} A_{43} A_{54} | B_i \rangle =
-[m_{15}+4][m_{55}][6]^{-1}[1-Y],
\eeq
\[
\langle B_i | A_{34} A_{45} A_{54} A_{43} | B_i \rangle
\]
$$
= \frac {[Y/2][Y/2+1]-[I][I+1]}{[2][3]}
\Bigl( \frac{[m_{15}+1][m_{55}-3]}{[3]}-
\frac { [2] [m_{15}+4][m_{55}] }{[6]} \Bigr)
$$
\beq
\hspace {9mm} -\frac { [Y/2-1][Y/2-2]-[I][I+1]}{[2][5]}
\biggl( [m_{15}-1][m_{55}-5]+\frac{[4]}{[6]}[m_{15}+4][m_{55}]\biggr),
\eeq
\beq
[2] \langle B_i | A_{54} A_{43} A_{34} A_{45}| B_i \rangle \sim
 [m_{15}-3][m_{55}-7]=[3-Y/2][2-Y/2]-[I][I+1].
\eeq
Exact coefficient of proportionality in the matrix element (10)
is unimportant, since it can be absorbed by redefining of $\beta.$

It is clearly seen from the definition of $q$-quantities
(see footnote 3) that baryon masses which follow from
(7)-(10) depend on hypercharge $Y$ and isospin $I$ (and,
hence, on $SU(3)$-breaking effects) in
highly nonlinear -- {\it nonpolynomial} -- fashion.

Substitution of (7)-(10) in (5),(6) and then in (1) gives
final expressions for $M_N$, $M_\Xi$, $M_\L$, $M_\S$.
Excluding from these the unknown constants $M_0,\a$ and $\b$,
we obtain the $q$-deformed mass relations of the form [6,7,13]
\[
[2]M_N+{[2]\over [2]-1}M_{\Xi }=[3]M_{\Lambda }
+\Bigl ({[2]^2\over [2]-1}-[3]\Bigr )M_{\Sigma }     \hspace{14mm}
\]
\vspace{-6mm}
\beq   \hspace{40mm}
  +{A_q\over B_q}\left (M_{\Xi } + [2] M_N -
               [2]M_{\Sigma } - M_{\Lambda } \right )\ ,
 \eeq
where $A_q$ and $B_q$ are certain polynomials of $[2]_q$ with
non-coinciding sets of zeros. It should be emphasized that
different dynamical representations, see the dependence on
$m_{15}$ and $m_{55}$ in (7)-(10), produce
different pairs $A_q$, $B_q$.
Any $A_q$ (rewritten in factorized form)
possesses the factor $([2]_q-2)$ and, thus, the "classical"
zero $q=1$.
In the limit $q=1$, each $q$-deformed mass relation reduces to
the standard GMO sum rule for octet baryons.
At some value(s) of $q$  which are zeros (other than
$q=1$) of particular $A_q$, we obtain mass sum rules which hold
with better accuracy than the GMO one.
The two mass sum rules
\beq
M_N+\frac{1+\sqrt{3}}{2}M_\Xi=\frac{2}{\sqrt{3}}M_\L+
\frac{9-\sqrt{3}}{6}M_\S\ ,
\eeq
\vspace{-4mm}
\beq
M_N+\frac{1}{[2]_{q_7}-1}M_{\Xi}=
\frac{1}{[2]_{q_7}-1}M_{\Lambda}+M_{\Sigma}\ ,
\eeq
where $[2]_{q_7}=2 \cos(\pi/7)$,
are obtained [6,7,13] from two different dynamical representations
$D^{(1)}$ and $D^{(2)}$
with corresponding polynomials $A_q^{(1)}$ and $A_q^{(2)}$,
by fixing zeros $q=\exp(\i\pi/6)$ and $q=\exp(\i\pi/7)$, respectively.
These sum rules show the precision of, resp., $0.22\%$ and $0.07\%$,
which is essentially better than the precision $0.58\%$ of GMO.
The case corresponding to $q=\exp(\i\pi/7)$ turns out [13] to be
the best possible one. It was supposed in [7] that this value
of $q$ may be interpreted as follows: $\pi/7 = 2 \theta_C$ (Cabibbo angle).

\bigskip
{\bf 3.\ }
Since our goal is to analyze a high nonlinearity in
$SU(3)$ breaking effectively accounted by the model,
let us first check the classical limit $q \to 1$ for the expressions
(7)-(10). These can be summarized to give the formula
\beq
M_{B_i}=M_{B(Y,I)}= M_0+ \alpha Y + \beta ( Y^2/4 - I ( I+1 ))
\eeq
for masses from the octet,
by suitable redefinition of $M_0,\alpha,\beta$. This relation
coincides with the mass formula of Gell-Mann and Okubo [9,11],
as it should be.

Now consider, at $q$ which is close to $1$ and taken to be pure
phase: $q=e^{ih}$, first few terms of the Tailor expansion for
the expressions (7)-(10) which enter baryon octet masses.
Using the formulas
\beq
[n]=\frac{\sin(nh)}{\sin(h)}=n-\frac{n(n^2-1)}{6}h^2+O(h^4),
\eeq
\vspace{-4mm}
\[
[n][n+1]=n(n+1)-h^2\left(\frac{(n(n+1))^2}{3}-\frac{n(n+1)}{6}\right)
+O(h^4)
\]
valid for small $h,$ from (7)-(10) we get
\[
M_{B_i}=M_0+\a\Bigl(-\bigl( (Y/2)(Y/2+1)-I(I+1)\bigr)
(10/3-161h^2+5h^2/9)
\]
\[
+\bigl((Y/2-1)(Y/2-2)-I(I+1)\bigr)(8-84h^2+4h^2/3)+
\]
\[
+\bigl((Y/2)^2(Y/2+1)^2-(I(I+1))^2\bigr)(10h^2/9)
\]
\[
-\bigl((Y/2-1)^2(Y/2-2)^2-(I(I+1))^2\bigr)(8h^2/3)\Bigr)
\]
\[
+\b\Bigl( \bigl((Y/2-3)(Y/2-2)-I(I+1)\bigr)(1+h^2/6)
\]
\beq
-\bigl((Y/2-3)^2(Y/2-2)^2-(I(I+1))^2\bigl)(h^2/3)\Bigr)\ ,
\eeq
where, for simplicity, the choice
$m_{15}=9,\ \ m_{55}=0$ has been fixed.

It is instructive to compare this result with the expansion in [9]
in terms of the $SU(3)$-breaking interaction, whose
lowest orders involve summands with the following
dependences on hypercharge and isospin:
\[ Y,\ \ \ \ (Y^2/4-I(I+1))  \ \ \ \ -\ \ \
\hbox {\it 1st order terms in\ }SU(3) \hbox{\it\ breaking}
\]
\[ Y^2,\ \ Y(Y^2/4-I(I+1)),\ \ (Y^2/4-I(I+1))^2 \ -\ \ \
\hbox {\it 2nd \ \ \ "\ \ \  "  }
\]
\[ Y^3,\ \ Y^2(Y^2/4-I(I+1)),\ \ Y(Y^2/4-I(I+1))^2,\ \ (Y^2/4-I(I+1))^3
   \  -\ \  \hbox {\it 3rd \ \ "\ \ "}
\]
\[ Y^4,\ \ Y^3(Y^2/4-I(I+1)),\ \ ... \ \ \ \ \ \ -\ \ \
\hbox {\it 4th \ \ \ "\ \ \ "}\ .
\]
Let us remark that the 1st order (the top row above)
is the highest possible one that allows getting a
mass sum rule {\it for octet baryons} in the traditional approach [9]
which treats the constants assigned to different terms
{\it as independent}.
On the other hand, dependence of baryon masses on $q$ through
the $q$-quantities like (15) determines unambiguously the coefficients
of expansions in $h$ and, thus, there appear no new parameters.
This, together with the fact that the deformation
parameter $q$ appears in (expressions for) baryon masses
through representation matrix elements  ({\it due to
deformation of symmetry}), shows a sharp distinction
of the dimensionless parameter $q$ from the
dimensionful constants $M_0,\a,\b$, introduced explicitly
in the mass operator (2) as symmetry  breaking parameters.

It is seen that the first order terms in $h^2$ in
the expression (16) correspond to terms up to the 4th order
of Okubo's expansion, some of which are
$(Y^2/4-I(I+1))^2,\ $ $Y^2(Y^2/4-I(I+1)),\ $ $Y^4$.
On the other hand, there is the 2nd order term in Okubo's expansion,
namely $Y(Y^2/4-I(I+1))$, which does not appear in the expression (16).
This means that the expansion in terms of small $h^2$ is consistent
with, but not the same as, the expansion in terms of $SU(3)$ breaking.

\bigskip
{\bf 4.\ }
Formula (14) is valid not only for octet baryons but also for the
$J^P=\frac32^+$ baryons from decuplet ${\bf 10}$ of $U_q(su_3)$.
Taking into account the specific property that, for the decuplet,
hypercharge and isospin obey the relation
\beq
I=1+Y/2\ ,
\eeq
one can rewrite (14) in the form $ M=M_0+\alpha Y$,
which produces the equal spacing rule for decuplet baryon masses.

It is easy to see that the expressions (7)-(10) which lead
in the $q$-deformed ($q\ne 1$) case to octet baryon masses
are equally well applicable in the decuplet case.
Indeed, using (17) and the easily verifiable identity
\[
[x-Y/2][x+1-Y/2]-[I][I+1]= - [Y-x+1][x+2]
\]
valid for all $x$,
we arrive at the $q$-average formula for masses
of decuplet baryons:
\beq
\frac{ M_{\Omega}-M_{\Xi^*} + M_{\Sigma^*}-M_{\Delta} }
      {[2] }
=M_{\Xi^*}-M_{\Sigma^*}  \ .
\eeq
This formula previously was obtained in [5] within
somewhat another context and shown to possess the important
property of {\it universality} (independence on the choice of a
dynamical representation of $U_q(su_5)$ or $U_q(su_{4,1})\ $ [5],
under the only condition that such a dynamical representation
contains the 20-plet $\{p+3,p,p,p\}$ of $U_q(su_4)$ in
which the $U_q(su_3)$-decuplet is embedded).
For example, within the dynamical irrep $\{p+4,p,p,p,p\}$ of
$U_q(su_5)$, the masses are  $M_\D=M_{10}+\b$,
$M_{\S^*}=M_{10}+\a+[2]\b$, $M_{\Xi^*}=M_{10}+[2]\a+[3]\b$,
$M_{\O}=M_{10}+[3]\a+[4]\b$, and these obviously satisfy (18).
Since each isoplet from the baryon decuplet is uniquely
fixed by its strangeness (or hypercharge) value, all these expressions
can be comprised by the single mass formula
\beq
M_{B_i^*}=M_{10}+\a [1-Y] +\b [2-Y] \ ,
\eeq
where $B_i^*$ runs over four different isoplets in ${\bf 10}$-plet.
From definition of $q$-numbers, it follows that
the dependence of both quantities: $[1-Y]$, $[2-Y]$ on hypercharge
$Y$ is essentially nonlinear and becomes linear only in the
classical (non-deformed) limit $q=1$.

Comparison of the relation (18) with empirical data for baryon
$J^P=\frac32^+$ masses [14] is successful if $q$ is fixed as
$q=exp(\i\theta_{\bf 10}),$ $\theta_{\bf 10}\simeq \pi/14$.
The latter angle, as argued in [7], can be juxtaposed with
the Cabibbo angle $\theta_C$.

\bigskip
{\bf 5.\ }
Up to now we used only a representation-theoretic part of
the structure of the quantum algebra $U_q(su_n)$. In this
section, we follow somewhat different approach and treat the mass
operator on the base of $q$-tensor operators. To this end, below
we will need such ingredients of the  Hopf algebra structure
of $U_q(su_n)$ as comultiplication $\D$ and
antipode $S$ operations.  These  are defined on the generators
$E_i^+\equiv A_{i,i+1}$,$E_i^-\equiv A_{i+1,i}$  and
$H_i\equiv A_{ii}-A_{i+1,i+1}$
according to the formulas [1]:
\[
\Delta (H_i)= H_i\otimes 1+1\otimes H_i\ \ ,\ \ \ \
S(H_i)=-H_i\ \ ,\ \ \
S(q^{H_i/2})=q^{-H_i/2}\ \ ,\ \ \ S(1)=1,
\]
\beq
\Delta(E_i^\pm)=E_i^\pm\otimes q^{H_i/2} +
q^{-H_i/2}\otimes E_i^\pm \ \ ,
\ \ \ \ S(E_i^\pm)=-q^{H_i/2}E_i^\pm q^{-H_i/2}.
\eeq
Consider the adjoint action of $U_q(su_n)$ defined as [12]
\[
ad_A B = \sum { A_{(1)} B S(A_{(2)})}\ ,
\]
where $A,B \in U_q(su_n),$
$A_{(1)}$ and $A_{(2)}$ are defined by
comultiplication
$ \Delta (A) = \sum { A_{(1)} \otimes A_{(2)}}.$
With the account of (20) this yields
\[
ad_{H_i} B=H_i B 1+1 B S(H_i)=H_i B-B H_i\ ,
\]
\[
ad_{E_i^\pm}B=
E_i^\pm B q^{-H_i/2}-q^{-H_i/2}B q^{H_i/2}E_i^\pm q^{-H_i/2}\ .
\]

Below we will need the $q$-tensor operators [15]
$(V_1,V_2,V_3)$ and $(V_{\bar{1}},V_{\bar{2}},V_{\bar{3}}),$
which transform under $U_q(su_3)$ as ${{\bf 3}}$
and ${{\bf 3}^*}$, respectively.
Let us denote $[X,Y]_q\equiv XY-qYX$. Direct calculations show
that the triple of elements from $U_q(su_4)$,
\[
 V_1=[E_1^+,[E_2^+,E_3^+]_q]_q q^{-H_1/3-H_2/6} \ ,\ \
V_2=[E_2^+,E_3^+]_q q^{H_1/6-H_2/6} \ ,\ \
\]
\beq
V_3=E_3^+ q^{H_1/6+H_2/3} \ ,
\eeq
transforms as ${{\bf 3}}$ under the adjoint action of $U_q(su_3).$
Moreover, $V_1$ corresponds to the highest weight vector, the pair
$(V_1, V_2)$ is an isodoublet and $V_3$ isosinglet under
$U_q(su_2).$


Likewise, by direct calculation it can be shown that the triple
of elements from $U_q(su_4)$,
\[
 V_{\bar{1}}=q^{-H_1/3-H_2/6} [E_1^-,[E_2^-,E_3^-]_{q^{-1}}]_{q^{-1}}
\ \ ,\ \
V_{\bar{2}}=q^{H_1/6-H_2/6} [E_2^-,E_3^-]_{q^{-1}} \ \ ,
\]
\beq
 V_{\bar{3}}=q^{H_1/6+H_2/3} E_3^- \ ,
\eeq
 transforms as ${{\bf 3}}^*$ under the adjoint action of
 $U_q(su_3).$ Moreover, $V_{\bar{3}}$ corresponds to the highest weight
vector, the pair $(V_{\bar{1}}, V_{\bar{2}})$ is an isodoublet and
$V_{\bar{3}}$ isosinglet under $U_q(su_2).$

As before, we take $U_q(su_3)$ broken down to $U_q(su_2)$ as the algebras
of global internal symmetry of hadrons
and make use of the correspondence
between baryons $J^P=\frac12^+$ and basis vectors in the
representation space of ${\bf 8}$ as well as between
baryons $J^P=\frac32^+$ and basis vectors
in the representation space of ${\bf 10}.$
Like in the case of the usual nondeformed flavor symmetry algebra
$su(3)$  broken down to its subalgebra $su(2)$,
we take the mass operator in the form
\beq
\hat M = \hat M_0+ \hat M_8 \ ,
\eeq
where $\hat{M}_0$ is a scalar of $U_q(su_3),$
$\hat{M}_8$ is the operator which transforms as the $I=0,Y=0$ component
of the tensor operator of ${\bf 8}$-irrep of $U_q(su_3).$

If $|B_i\rangle$ is a basis vector of the representation
${\bf 8}$ (or ${\bf 10}$) space which corresponds to some
baryon with spin $J=1/2$ (or $J=3/2$), respectively, then
$M_{B_i}=\la B_i|\hat{M}|B_i \ra$ is the mass of this baryon, see (1).

Consider first the case of octet baryons. The irrep
${\bf 8}$ occurs twice in the decomposition
\beq               \qquad \qquad \qquad
{{\bf 8}} \otimes {{\bf 8}} =
{{\bf 1}} \oplus {{\bf 8}}^{(1)} \oplus
{{\bf 8}}^{(2)} \oplus
{{\bf 10}}^* \oplus {{\bf 10}} \oplus
{{\bf 27}}\ .
\eeq
This fact, the Wigner-Eckart theorem for $U_q(su_n)$ quantum algebras [15]
applied to $q$-tensor operators transforming as irrep ${\bf 8}$
of $U_q(su_3)$, and symmetry properties [16] of $q$-Clebsch--
Gordan coefficients lead us to the conclusion that the mass operator is
of the form
$$              \hspace{-94mm}
\hat M = M_0 {\bf 1} + \a V_8^{(1)} + \b V_8^{(2)}
               \eqno{(23')}
$$
and the baryon masses are calculated as
\beq
M_{B_i} =
\langle B_i |(\hat M_0+ \hat M_8) | B_i \rangle =
\langle B_i |(M_0 {\bf 1} + \a V_8^{(1)} + \b V_8^{(2)})
| B_i \rangle\ .
\eeq
Here  ${\bf 1}$ is the identity operator,
$V_8^{(1)}$ and $V_8^{(2)}$ are two fixed tensor operators
with non-proportional matrix elements,
which both have the same transformation property as
the $I=0, Y=0$  component of  irrep ${\bf 8}$ of $U_q(su_3)$
(i.e., the same as that of $\hat {M_8}$);
$M_0,$ $\a$ and $\b$ are some unknown constants
depending on details (dynamics) of the model.

In the decuplet case, operator ($23'$) is equivalent to the operator
\[
\hat M = M_0 {\bf 1} + \tilde \a V_8 \ .
\]
This follows from the fact that the irrep ${{\bf 8}}$
occurs only once in  the decomposition
$$
{{\bf 10}}^* \otimes {{\bf 10}} =
{{\bf 1}} \oplus {{\bf 8}} \oplus
{{\bf 27}} \oplus {{\bf 64}} \ .   \qquad \qquad \qquad \qquad\qquad
$$
Formally, we may use the operator ($23'$) in the case
of decuplet baryons, too. But in this case, the  matrix elements
of $V_8^{(1)}$ and $V_8^{(2)}$ become proportional to each other,
and that effectively leads to a single constant $\tilde \a$
instead of the two $\a$ and $\b$ in ($23'$).

From the decompositions
\beq        \qquad\qquad \quad
{{\bf 3}} \otimes {{\bf 3}}^* =
{{\bf 1}} \oplus {{\bf 8}},\ \ \ \ \
{{\bf 3}}^* \otimes {{\bf 3}} =
{{\bf 1}} \oplus {{\bf 8}}  \ ,              
\eeq
it is seen that the operators $V_{3}\ V_{\bar{3}}$
and $V_{\bar{3}}\ V_{3}$ from (21),(22) are
just the two isosinglets needed in the context of equation ($23'$).
It follows from decompositions (26) that each of them transforms
as the sum of a singlet (scalar) and the $I=0, Y=0$ component
of the octet.  Hence, the mass operator ($23'$) can be rewritten
(after redefinition of $M_0,\a,\b$) in the equivalent form
\[
\hat M = M_0 {\bf 1} +\a V_3 V_{\bar{3}}
+ \b  V_{\bar{3}} V_3
\]
or
 \vspace{-2mm}
\beq
\hat M = M_0 {\bf 1} +\a E_3^+ E_3^- q^Y + \b E_3^- E_3^+ q^Y \ ,
\eeq
where the formula $Y=(H_1+2H_2)/3$ for hypercharge
is used.

To obtain matrix elements
(25), we use an embedding of ${\bf 8}$ or ${\bf 10}$
into some concrete representation of $U_q(su_4).$
Embedding the octet ${\bf 8}$ of $U_q(su_3)$, for instance,
into ${\bf 15}$ (adjoint representation) of $U_q(su_4)$,
on the base of (25),(27) we obtain the following expressions
for the octet baryon masses:
\beq
 M_N=M_0+\b q \ \ ,\ \ \ M_\S= M_0\ \ ,
\ \ M_\Lambda =M_0 + \frac{[2]}{[3]} (\a+\b)\ \ ,
\ \ M_\Xi=M_0+\a q^{-1}\ . \
\eeq
Let us emphasize that the expressions for $M_N, M_\Xi$ are
not invariant under $q \to q^{-1}$ and no transformation
 of $M_0,\a,\b$ exists which makes them invariant.

Excluding  $M_0,\a$ and $\b$ from (28), we obtain
the following $q$-analogue of the GMO formula for octet baryons:
\beq
[3]M_\Lambda + M_\S=[2](q^{-1} M_N+q M_\Xi)\ .
\eeq
Observe apparent simplicity of (29) as compared with $q$-MR (11).
This same formula (29) is obtained by embedding
${\bf 8}$
into other dynamical representations of $U_q(su_4).$

However, what concerns validity of (29)
with empirical baryon masses [14],
there is no other way to fix the deformation parameter $q$
than to apply a fitting procedure. One can check that for
each of the values $q_{1,2}=\pm 1.035$,
$q_{3,4}=\pm 0.903 \sqrt{-1}$, the left hand side of $q$-MR (29)
coincides with its r.h.s within experimental uncertainty
(note that for $q_3,\ q_4$ the constants $\a$ and $\b$ in (28)
must be pure imaginary).
This is in sharp contrast with the $q$-MR (11), for which there
exists an appealing possibility to fix the parameter $q$
in a rigid way by taking zeros of relevant polynomial $A_q$,
see the discussion after (11) as well as [6,7].

\bigskip
{\bf 6.\ }
It is clear that the r.h.s. of (29) is invariant
under $q \to q^{-1}$ only if $q=q^{-1}$, that is, $q=\pm 1.$
Here we make an interesting observation.

Behind the "classical" mass formula of Gell-Mann and Okubo
which obviously follows from (29) at $q=1$ and
corresponds to the nondeformed unitary symmetries
$SU(4) \supset SU(3) \supset SU(2)$, there is also an unusual
"hidden symmetry" reflecting the singular $q=-1$ case of
$U_q(su_4) \supset U_q(su_3) \supset U_q(su_2)$ algebras
(undefined in this case). However, the relevant
objects exist as certain operator algebras. Let us describe
them in the part corresponding to $n=2$ and $n=3$.

At generic $q$, $q\ne -1$, the algebra $U_q(su_2)$ is generated by the
elements $E_1^+$,$E_1^-$ and $H_1$, which satisfy the relations
(here $[A]$ denotes $(q^A-q^{-A})/(q-q^{-1})$):
\beq
[H_1,E_1^\pm]=\pm 2 E_1^\pm ,\ \ \ [E_1^+,E_1^-]=[H_1] .
\eeq
In the limit $q\to 1$ it reduces to the classical algebra $su_2$.
We take the representation spaces of the latter in order to
construct operator algebras for the case $q=-1$. To each  $su_2$
representation space given by $j$ (which takes integral or
half-integral nonnegative values) with basis elements
$|j m\ra,\ m=-j,-j+1,\ldots,j$,
there corresponds a concrete operator algebra
generated by the operators defined according to the formulas
\[
H_1|j m\ra=2m|j m\ra ,\ \ \
E_1^+|j m\ra=\a_{j,m}|j m+1\ra ,\ \ \
E_1^-|j m\ra=\a_{j,m-1}|j m-1\ra\ \ \
\]
where
\[       \vspace{-3mm}
\a_{j,m}=\cases{\sqrt{-(j-m)(j+m+1)},\ \ \ \
                             j \ \ \ {\hbox{is an integer}},
\cr{}\cr\sqrt{(j-m)(j+m+1)},\ \ \ \  \
                        j \ \ \ {\hbox{is a half-integer}} .\cr}
      \vspace{1mm}    \]
The so defined operators
$E_1^+$,$E_1^-$ and $H_1$ on the basis elements
$|j m\ra$ satisfy the relations (compare with (30)):
\[            \vspace{-3mm}
[H_1,E_1^\pm]=\pm 2 E_1^\pm,\ \ \
[E_1^+,E_1^-]=\cases{ -H_1,\ \ \ \ j \ \ \ \hbox{is an integer};
\cr {} \cr H_1,\ \ \ \ \ j \ \ \ \hbox{is a half-integer} .\cr}
\]

In order to consider the (singular) case $q=-1$ of $U_q(su_3)$,
it is more convenient to deal with $U_q(u_3)$. We take
a representation space $V_\chi$, labelled by
$\{m_{13},m_{23},m_{33}\}\equiv\chi$, of the nondeformed $u_3$
and the Gel'fand-Tsetlin basis with elements
$|\chi;m_{12},m_{22};m_{11}\ra$ in each $V_\chi$.
Define the operators $E_1^+$,$E_1^-$,$H_1$,
$E_2^+$,$E_2^-$, $H_2$ that form the operator algebra
of the $\chi$-type by their action according to the formulas
\[
H_2|\chi;m_{12},m_{22};m_{11}\ra=
(2m_{12}+2m_{22}-m_{13}-m_{23}-m_{33}-m_{11})
|\chi;m_{12},m_{22};m_{11}\ra \ ,
\]
\[
E_2^+|\chi;m_{12},m_{22};m_{11}\ra=
\]
\[
a_{\chi,m_{11}}(m_{12},m_{22})
|\chi;m_{12}+1,m_{22};m_{11}\ra  + b_{\chi,m_{11}}(m_{12},m_{22})
|\chi;m_{12},m_{22}+1;m_{11}\ra \ ,
\]
\[
E_2^-|\chi;m_{12},m_{22};m_{11}\ra=
\]
\[
a_{\chi,m_{11}}(m_{12}-1,m_{22})
|\chi;m_{12}-1,m_{22};m_{11}\ra  + b_{\chi,m_{11}}(m_{12},m_{22}-1)
|\chi;m_{12},m_{22}-1;m_{11}\ra \ ,
\]
where
\[
a_{\chi,m_{11}}(m_{12},m_{22})=
\]
\[
\left(
(-1)^{m_{11}+m_{13}+m_{23}+m_{33}}
\frac{(m_{13}-m_{12})(m_{23}-m_{12}-1)(m_{33}-m_{12}-2)(m_{11}-m_{12}-1)}
{(m_{22}-m_{12}-1)(m_{22}-m_{12}-2)}
\right)^{1/2} \ ,
\]
\[
b_{\chi,m_{11}}(m_{12},m_{22})=
\]
\[
\left(
(-1)^{m_{11}+m_{13}+m_{23}+m_{33}}
\frac{(m_{13}-m_{22}+1)(m_{23}-m_{22})(m_{33}-m_{22}-1)(m_{11}-m_{22})}
{(m_{12}-m_{22}+1)(m_{12}-m_{22})}
\right)^{1/2}\ .
\]
Action formulas for the operators $E_1^{\pm}$ and $H_1$ are completely
analogous to the formulas given above for $n=2$ (with account of
$m_{11}-m_{22}= 2 j ,\ 2m_{11}-m_{12}-m_{22}= 2 m$).

The presented action formulas for the operators that form the
operator algebra of the $\chi$-type show that their matrix elements are,
to some extent, similar to the "classical" matrix elements
(i.e., to the matrix elements of the irrep $\chi$ operators
for $su(n)$).
However, there is an essential distinction: now we observe
the important phase factors
(namely, $(-1)^{m_{11}+m_{13}+m_{23}+m_{33}}$
under the square root in $a_{\chi,m_{11}}$ and $b_{\chi,m_{11}}$)
which depend on $\chi$ and a concrete  basis  element.
No such {\it basis-element dependent} factors exist in the $su(n)$ case.

Let us illustrate the treatment with the example of the operator
algebra which replaces the singular (undefined)
$q=-1$ case of $U_q(su_3)$ and
{\it corresponds to octet representation of $su_3$}.
We give explicitly those action formulas for $E_1^\pm$ and $E_2^\pm$,
in which matrix elements differ from their corresponding
"classical" counterparts:
\[
E_1^-|\S^+\ra=\sqrt{-2}|\S^0\ra,\ \ \
E_1^-|\S^0\ra=\sqrt{-2}|\S^-\ra,\ \ \
\]
\[
E_1^+|\S^-\ra=\sqrt{-2}|\S^0\ra,\ \ \
E_1^+|\S^0\ra=\sqrt{-2}|\S^+\ra,\ \ \
\]
\[
E_2^-|n\ra=\frac{1}{\sqrt{-2}}|\S^0\ra+
\sqrt{-\frac{3}{2}}|\L\ra,\ \ \
E_2^-|\L\ra=\sqrt{-\frac{3}{2}}|\Xi^0\ra,\ \ \
E_2^-|\S^0\ra=\frac{1}{\sqrt{-2}}|\Xi^0\ra,
\]
\[
E_2^+|\Xi^0\ra=\frac{1}{\sqrt{-2}}|\S^0\ra+
\sqrt{-\frac{3}{2}}|\L\ra,\ \ \
E_2^+|\L\ra=\sqrt{-\frac{3}{2}}|n\ra,\ \ \
E_2^+|\S^0\ra=\frac{1}{\sqrt{-2}}|n\ra.
\]
To complete this operator algebra, we must add the rest of action formulas
for $E_1^\pm$ and $E_2^\pm$ (i.e., action on those basis elements),
which coincide with "classical" ones,
as well as the action formulas for $H_1$, $H_2$
(these latter also coincide with "classical" formulas).

Analogously, proper operator algebras can be given which
correspond to any other irrep of $su_3$.

\bigskip
{\bf 7.\ }
We have demonstrated that applying, instead of customary hadronic
flavor symmetries, the corresponding quantum algebras to
derivation of baryon mass formulas takes effectively into account
a high nonlinearity (even nonpolynomiality) of baryon masses in
$SU(3)$ breaking or, equivalently, in
hypercharge $Y$ and isospin $I$. This is clearly
exhibited by formulas (7)-(10), (15) in the octet case and
(19), (15) in the decuplet case. Using techniques of
$q$-tensor operators, we obtained the new version (29) of $q$-MR.
Unlike the previously obtained $q$-MRs (11), this version
does not respect the symmetry under $q \to q^{-1}$.
We have found that, besides $q=1$,
the case $q=-1$ also yields the standard GMO sum rule,
requiring however a special treatment. The corresponding
mathematical structure is supplied by operator algebras, as
given in Sect.6.

\bigskip
{\bf Acknowledgements.} We are thankful to Prof. A.Klimyk
for valuable remarks. The research described in this
publication was made possible in part by the Award
No. UP1-309 of the U.S. Civilian Research \& Development
Foundation for the Independent States of the Former Soviet
Union (CRDF), and by the Grant INTAS-93-1038-Ext.

\end{document}